\documentclass[prb,twocolumn,floatfix]{revtex4-2}
\usepackage{float}
\usepackage[T1]{fontenc}
\usepackage{amsmath}
\usepackage{textcomp}
\usepackage{graphicx,bm}
\usepackage{blindtext, rotating}
\usepackage{mathtools}
\usepackage{gensymb}
\usepackage{enumitem}
\usepackage{xcolor}
\usepackage{hyperref}
\usepackage{braket}
\usepackage{natbib}
\usepackage{tabularx}
\usepackage{amssymb}
\usepackage{verbatim}
\makeatother

\begin{document}
\title{Rashba splitting in polar-nonpolar sandwich heterostructure : A DFT Study}     
\author{Sanchari Bhattacharya}
\email{bh.sanchari@gmail.com}
\affiliation{Department of Physics and Astronomy, National Institute of Technology, Rourkela, Odisha, India, 769008}
\author{Sanjoy Datta}
\email{dattas@nitrkl.ac.in}
\affiliation{Department of Physics and Astronomy, National Institute of Technology, Rourkela, Odisha, India, 769008}
\date{\today}

\begin{abstract} 

In this study, we employ density functional theory (DFT) based first-principles calculations to investigate the spin-orbit effects in the electronic structure of a polar-nonpolar sandwich heterostructure namely LAO$_{2.5}$/STO$_{5.5}$/LAO$_{2.5}$. Our focus on the Ti-3d bands reveals an inverted ordering of the STO-$\rm t_{2g}$ orbital near the n-type interface, consistent with earlier experimental work. In contrast, toward the p-type interface, the orbital ordering aligns with the natural ordering of STO orbitals, influenced by crystal field splitting. Interestingly, we have found a strong inter-orbital coupling between $t_{2g}$ and $e_g$ orbital, which has not been reported earlier in $\rm SrTiO_3$ based 2D system. Additionally, our observations highlight that the cubic Rashba splitting in this system surpasses the linear Rashba splitting, contrary to experimental findings. This comprehensive analysis contributes to a refined understanding of the role of orbital mixing in Rashba splitting in the sandwich oxide heterostructures.
\end{abstract}
\maketitle

\section{Introduction}\label{introduction}
The pursuit of effective methods for generating and regulating spin-polarized electric currents is a paramount objective within the realm of spintronics, a field devoted to creating resilient, energy-efficient electronic devices that demonstrate non-volatile behaviour~\cite{wolf2001spintronics}. Fundamental breakthroughs in understanding spin-dependent phenomena in bulk semiconductors have been achieved through the exploration of Rashba spin-orbital coupling (RSOC)~\cite{rashba1961combinational,rashba1991landau, rashba1960properties, vas1979spin, rashba2015symmetry}. In a solid, the Rashba effect emerges from the interplay of intrinsic spin-orbit coupling (SOC) and the disruption of structural inversion symmetry. This interplay locks the spin and momentum degrees of freedom, leading to the elimination of spin degeneracy in the bands~\cite{knap-wal-qw,winkler2003spin,winkler-electron-hole,faniel-soc-qw}.
Moreover, the mixing of orbitals also plays a crucial role in determining the nature and the magnitude of Rashba spin-splitting (RSS) along with the strength of SOC in the system~\cite{varotto2022direct}. This aspect opens up the realm of spin-orbitronics.~\cite{bihlmayer-spin-orbitronics,sto-kto-spin-orb}.

The nature of RSOC makes it possible to control and tune the electron spin with only an external electric field~\cite{orbital-elec-spin}. 
Furthermore, it has the advantage of an improved interconversion between spin and charge currents~\cite{vaz2019mapping}, exhibiting phenomena like spin Hall effect~\cite{kato2004coherent,sinova-spin-hall}, Edelstein and inverse Edelstein effects~\cite{edelstein1990spin}. 
These properties have a significant impact on the advancement of quantum computers based on semiconductor quantum dots~\cite{vandersypen2019quantum} and have left an indelible mark on various domains within condensed matter physics~\cite{manchon2015new}. The field of spin-orbitronics started with semiconductors~\cite{dyakonov1971current} and metals~\cite{sanchez2013spin}.
Later on it expanded its domain in the non-centrosymmetric bulk oxides, such as $\rm BiAlO_3$~\cite{BAO-picozzi-2016}, 
$\rm KTaO_3$~\cite{venditti2023anisotropic}, $\rm KIO_3$~\cite{sheoran2022rashba}, $\rm PbTiO_3$~\cite{arras-pbtio3-soc}. 
Although the evidence of RSOC has been found in these bulk systems, the inherent requirement of ferroelectricity severely restricts the number of such materials. 
Moreover, a bulk system is not very suitable from the point of view of the fabrication of modern devices. Oxide-heterostructures, on the other hand, provide a much more flexible platform to explore the RSOC along with a plethora of unique features~\cite{soumyanarayanan2016emergent,bibes2011ultrathin,chen2023spintronics}.
The long spin lifetimes and modest RSOC seen on oxide surfaces and interfaces render them appealing candidates for spin-orbitronic devices~\cite{varignon2018new}. 
In particular, $\rm{SrTiO_3}$ have become one of the most widely studied materials after the detection of two-dimensional electron gas (2DEG) in its surface~\cite{sto-surface-2deg} and interface with $\rm{LaAlO_3}$ (LAO)~\cite{ohtomo2004high}.
2DEG could also be found in a slab~\cite{sto-slab-111}, thin film~\cite{sto-thin-film-so}, or at the surface~\cite{rebec2019dichotomy} of $\rm SrTiO_{3}$. This 2DEG could be metallic~\cite{metallic-sto-2deg} or semiconducting~\cite{ohtomo2004high} in nature, such as forming n-type and p-type interfaces~\cite{ohtomo2004high,pentcheva_2010}.
The bulk strontium titanate (STO) exhibits octahedral symmetry~\cite{piskunov2004bulk} with a band gap of $\approx 3.75~\rm{eV}$~\cite{van2001bulk}. The presence of time-reversal and inversion symmetries leads to a two-fold spin degenerate band structure in the bulk STO. This doubly degenerate nature of the STO band structure remains unaltered in the presence of intrinsic spin-orbit coupling. However, the introduction of a surface~\cite{santander2011-SrTiO3} or interface with other transition metal oxides~\cite{mannhart2010oxide,sto-lmo-hetero,li2017interfacial,wang2023high} disrupts inversion symmetry, leading to the lifting of this spin degeneracy. 
The internal electric field in this set of structures can be tuned by adjusting the slab thickness or by engineering the surface and the interface. The tuning of the internal electric field can also be achieved by another alternative method, in which an STO slab could be sandwiched between a polar material on either side that is arranged in a manner to generate
the internal electric field in the nonpolar STO.

A recent experimental study suggested that the manipulation of the internal electric field can be achieved by creating two oppositely 
charged interfaces in LAO/STO/LAO oxide heterostructures~\cite{lin2019interface}. This internal electric field effectively breaks the structural inversion symmetry in STO, leading to the emergence of RSOC.
Furthermore, it is claimed that by reducing the STO (0 0 1) slab thickness from 20 u.c to 8 u.c, the RSOC transitioned from a cubic to a linear type. 
This experimental work also concluded that the RSOC is of linear type in thin heterostructures in contrast to the previously reported cubic RSS in oxide interfaces. 
Their conclusion is based on the study of systems down to 8 u.c thin STO slab. However, there is no detailed comparison of experimental results with the first-principle study of the nature RSS in such a thin STO slab.

In this article, we conduct a thorough density functional theory (DFT) based first-principle analysis of the nature of Rashba coupling in the LAO$_{2.5}$/STO$_{5.5}$ (0 0 1)/LAO$_{2.5}$ structure. 
In particular, we focus on the electronic structure, with an emphasis on the role of orbitals and the spin texture associated with the spin-split bands.
Furthermore, we use an effective $\vec{k}.\vec{p}$ Hamiltonian to determine the RSOC coefficient and ascertain the nature of the splitting. We find that the experimental conclusion regarding the orbital ordering is in tune with the DFT results, while the nature of Rashba coupling stands in contrast  with our analysis.
\section{Computational methods}\label{Computational Details}
First-principle density functional theory computations are done with the Quantum Espresso (QE) package (QE)~\cite{qe,giannozzi2017advanced}. 
The Perdew-Burke-Ernzerhof (PBE)~\cite{pbe}~functional is used to account for electron exchange-correlation effects. Projector augmented wave (PAW)~\cite{paw}~basis set is used to account for valence electron-core ion interaction with a 75 Ry cut-off energy and 750 Ry cut-off charge density. All ionic relaxation are performed for geometry optimization using the quasi-Newton Broyden-Fletcher-Goldfarb-Shann (BFGS) algorithm~\cite{bfgs}. This study uses $\Gamma$-centered $8 \times 8 \times 1$ k-point grids, as suggested by Monkhorst and Pack~\cite{monkhorst1976special}, to calculate the first Brillouin zone.  
Using a minimum thickness of $\geq~30$~\AA~ vacuum layer in the z-direction (perpendicular to the plane) reduces interaction between adjacent slabs. The total energy convergence criteria for geometrical relaxation are set to be < $\rm 10^{-4}~Ry$. All the atomic positions are optimized until the interatomic forces are less than $\rm 10^{-3}~Ry/Bohr$.
The correlated $\rm Ti-3d$ state has been treated with ortho-atomic Hubbard correction using $\rm U=5~eV$ and $\rm J= 0.64~eV$ as reported in the earlier studies~\cite{Ti-hub-1,Ti-hub-2,Ti-hub-3,kong2021tunable}.
To examine collinear magnetism, spin-polarized computations were done on the slab system. Collinear magnetism was found to be missing from the system. Therefore, the slab system with a starting magnetism of $0~\mu_B$ on Ti was used for all subsequent calculations involving spin-orbit interaction.
\section{Structural details}
\begin{figure}[htbp!]
\includegraphics[width=0.49\textwidth,clip=true]{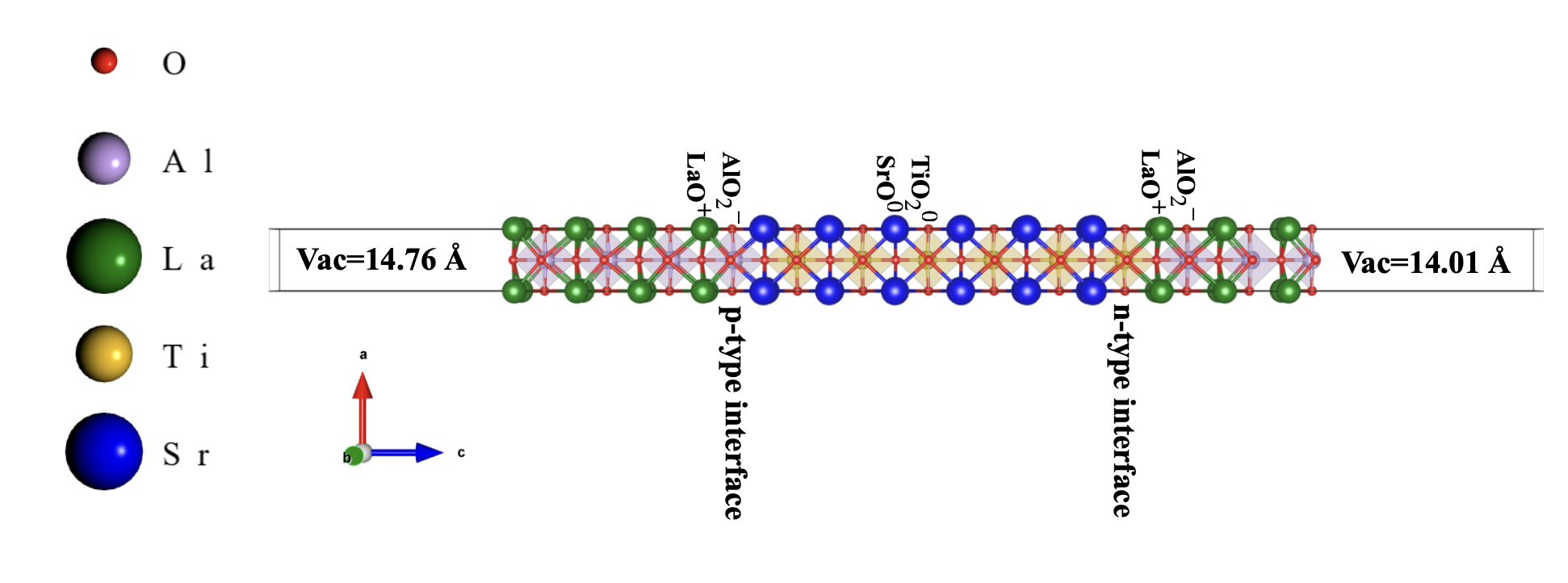}
\caption{Structural model for the LAO/STO/LAO sandwich hetero structure with n-type, i.e., $\rm {LaO}^{+}$/$\text{AlO}_\text{2}^{-}$ interface and 
p-type, i.e., $\text{AlO}_\text{2}^{-}$/$\text{SrO}^{0}$ at both sides of 5.5 u.c STO slab.}
\label{fig:pn-struct}
\end{figure}
In this study, we investigate the non-stoichiometric LAO/STO/LAO sandwich structure, where two unequal interfaces have been created to achieve 
tunable electric polarization as suggested in Ref.~\cite{lin2019interface}.
STO and LAO are both cubic perovskites with a general configuration of $\rm ABO_3$. LAO is composed of a stacked 
arrangement of $\rm {LaO}^{+}$ and $\text{AlO}_\text{2}^{-}$, whereas, STO are composed with the alternative arrangements 
of $\text{TiO}_\text{2}^{0}$ and $\text{SrO}^{0}$ layers. Eventually, to keep the stoichiometry intact, two types of interfaces
can be formed with these two materials, such as $\text{TiO}_\text{2}^{0}$ and $\rm {LaO}^{+}$ forming n-type interface and 
$\text{AlO}_\text{2}^{-}$ and $\text{SrO}^{0}$, forming p-type interface, as mentioned in the Section~\ref{introduction}.
In the present work, the theoretical simulation is done with a 5.5 uc slab of STO that has been put on the $\text{AlO}_\text{2}^{-}$-ended LAO substrate 
and this $\text{TiO}_\text{2}^{0}$-ended STO slab is capped with 2.5 u.c. LAO. In this particular arrangement, the disparity in potential between the two surfaces generates an inherent electric field that disrupts the inversion symmetry of STO and generates an internal electric field within the STO slab, as presented in Fig.~\ref{fig:pn-struct}.
The in-plane lattice parameter of the asymmetric sandwich structure is fixed as the theoretical lattice parameter of LAO, i.e., 3.813~\AA. 
The theoretical lattice parameter of STO is 3.939~\AA, so making the sandwich heterostructure implies a compressive strain on the 
STO $\approx$ -3.2 \%. In this structural geometry, the STO has a polarization along [0 0 1] with Sr and Ti ions below the oxygen atoms in each SrO and $\rm TiO_2$ layers.

\section{Results}
\subsection{Orbital ordering}
\begin{figure}[htbp!]
\includegraphics[width=0.5\textwidth,trim={1.2cm 3.0cm 1cm 3.2cm},clip=true]{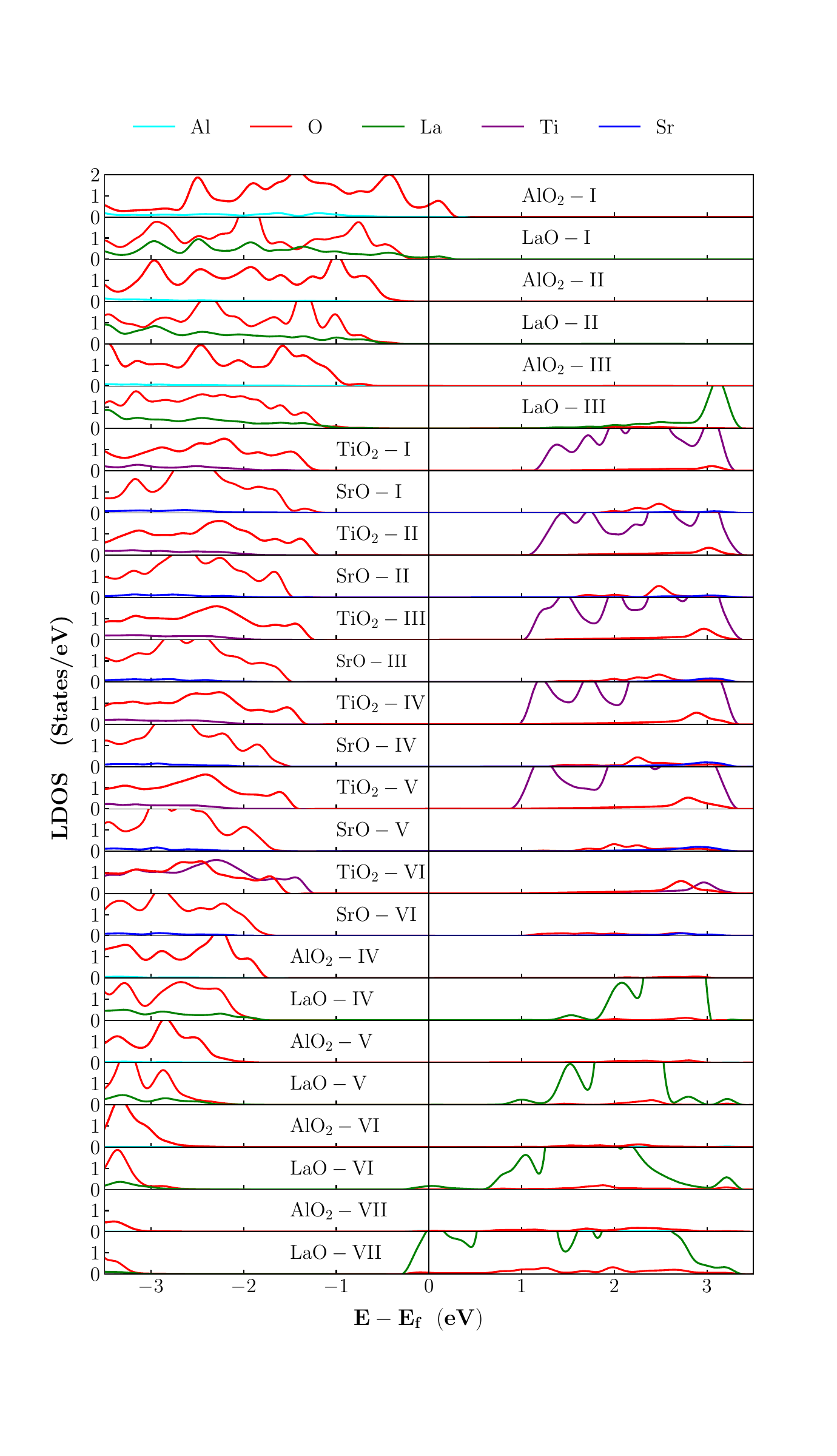}
\caption{Calculated layer-resolved density of states available for the LAO/STO/LAO asymmetric sandwich structure.}
\label{fig:lrdos}
\end{figure}
\begin{figure*}[htbp!]
\centering
\includegraphics[width=\textwidth]{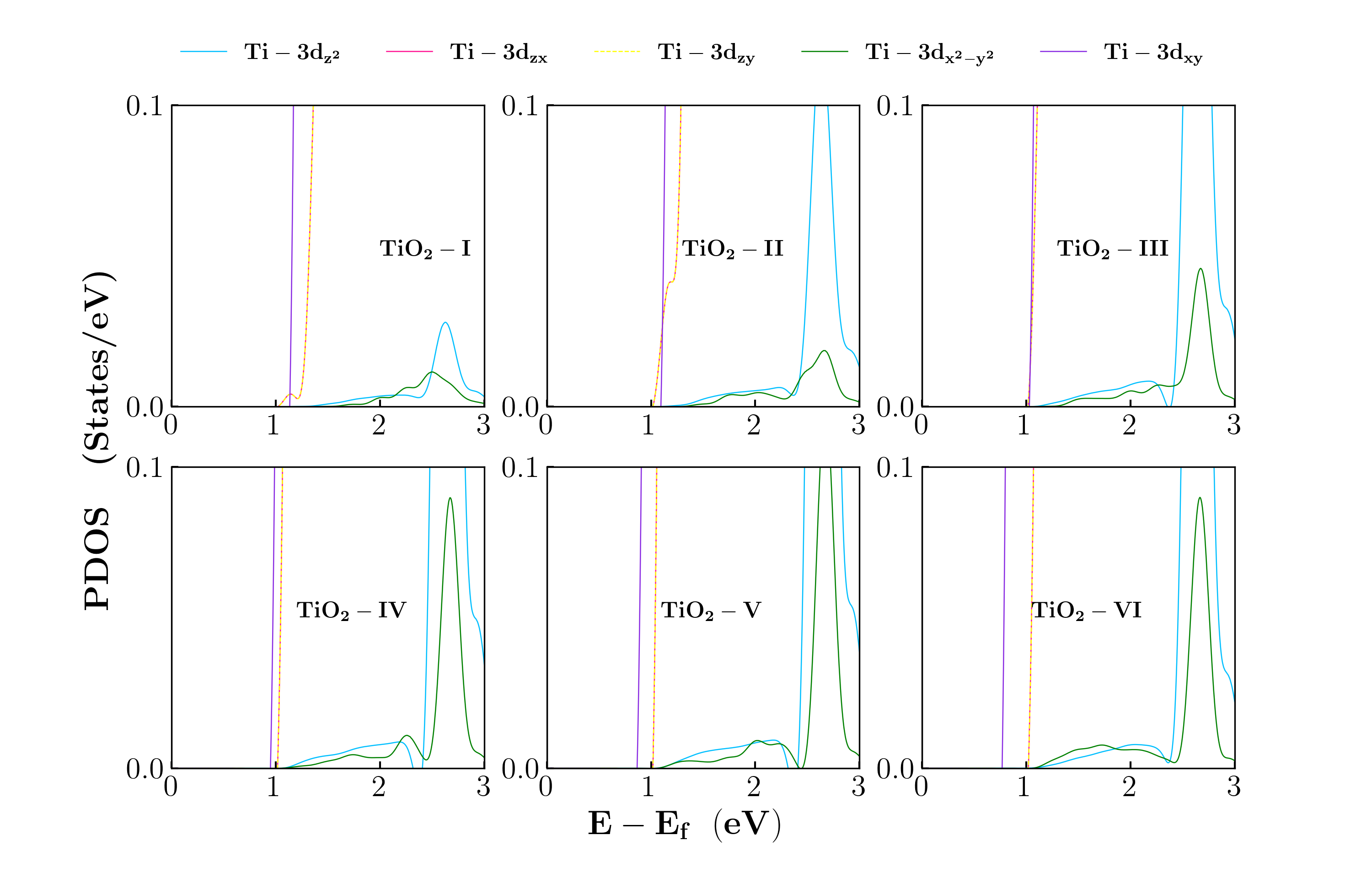}
\caption{Orbital resolved projected density of states for every $\rm{TiO_2}$ layer in the sandwich heterostructure. The shifting of orbital ordering is presented by the available states near the Fermi energy.}
\label{fig:orb-band}
\end{figure*}
We explore the stoichiometric LAO/STO/LAO sandwich heterostructure in Fig.~\ref{fig:pn-struct} in the presence of SOC. 
It is important to note that the electronic structure calculations reveal the system to be non-magnetic.
In Fig.~\ref{fig:lrdos}, we present the atomic layer-resolved total density of states (LRDOS) for each layer in the heterostructure. This figure indicates that due to the interface formation, the bands of the STO move towards the Fermi energy; however, there is no charge accumulation at the interface in the system. Moreover,  no metallic state is found throughout the STO slab. 
In previously reported STO-based heterostructures, the RSS is notably observed in the Ti-3d orbitals~\cite{sto-dband-rso}. Consequently, our study concentrates on the Ti-3d orbitals within this system. In the current structure, in the absence of SOC, Ti-3d orbitals are split into $\rm t_{2g}$ and $\rm e_g$ orbitals due to crystal field splitting. Due to slab formation, the $\rm t_{2g}$ levels are further splitted into $\rm d_{xy}$ and $\rm d_{xz/yz}$.
For direct visualization of the $\rm t_{2g}$ and $\rm e_g$ orbitals, we display the corresponding projected density of states for all $\rm TiO_2$ layers in Fig.~\ref{fig:orb-band}. This orbital-resolved representation offers insights into the orbital ordering of the Ti-3d bands originating from each layer.
Close to the n-type interface, in the $\rm{TiO_2}$-I and $\rm{TiO_2}$-II, the observed inverted orbital ordering of the $\rm t_{2g}$ orbital aligns with previously reported experimental findings~\cite{lin2019interface}. Interestingly, this inverted orbital ordering of $\rm t_{2g}$ orbital is specific in the vicinity of n-type interface. However, the orbital ordering changes as it moves towards the p-type interface ($\rm{TiO_2}$-III to $\rm{TiO_2}$-VI), which agrees with the reported orbital ordering of the $\rm{SrTiO_3}$ slab~\cite{sto-thin-film-so,shanavas-2016-sto}.
Interestingly, in contrast to the earlier report~\cite{lin2019interface}, when SOC is applied, the inverted orbital ordering is only found in the $\rm{TiO_2}$-I layer, and the remaining layers show the usual orbital ordering present in other STO-based interfaces and heterostructures.
\begin{figure}[htbp!]
\includegraphics[scale=0.5,trim={0.9cm 1.4cm 0.2cm 1.2cm},clip=true]{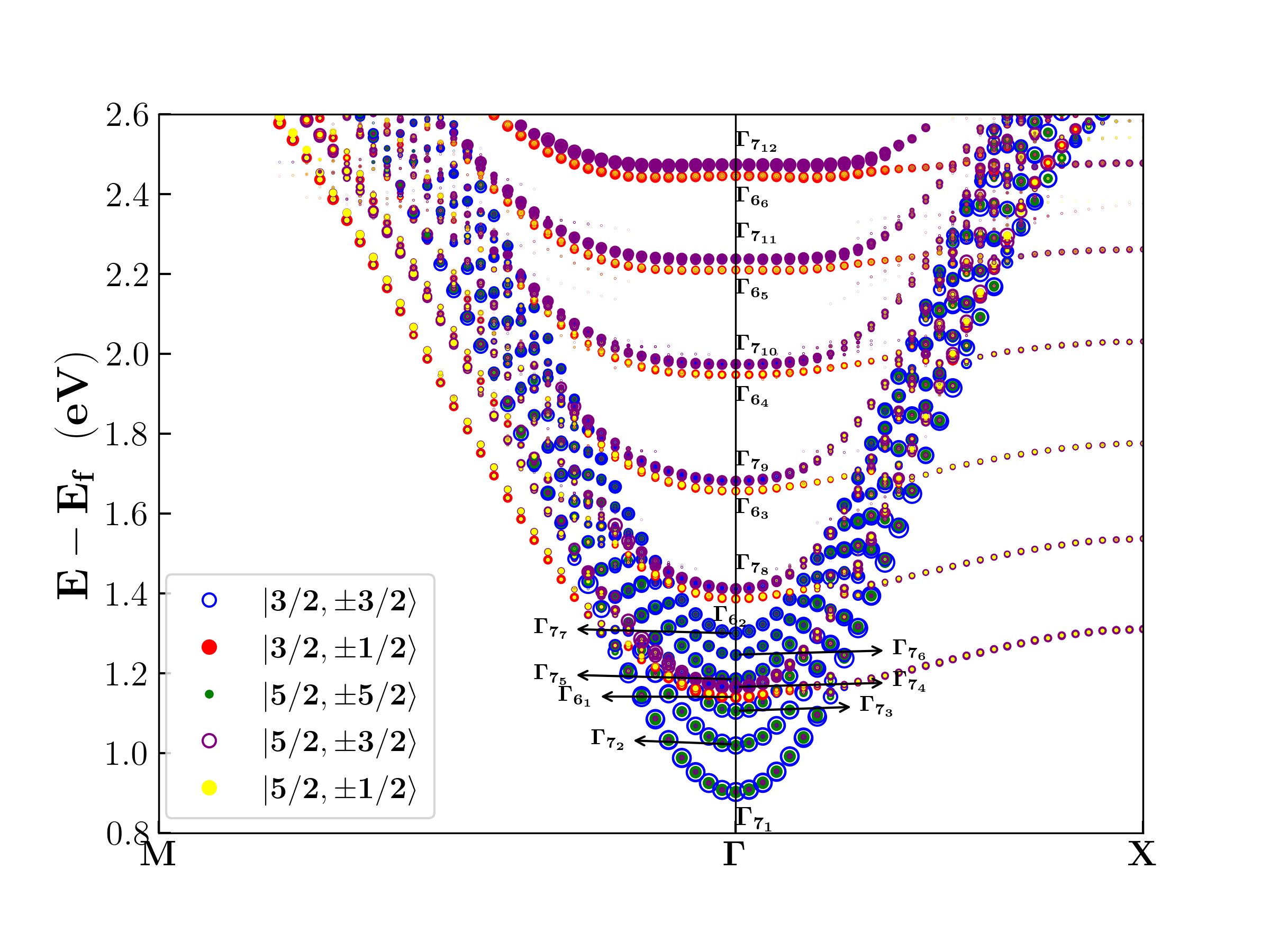}
\caption{Calculated fat band structure with the orbital angular momentum associated with the bands in the $\rm{SrTiO_3}$ layers. The bands are designated with their level as $\Gamma_{7_i}$ and $\Gamma_{6_i}$, as they splitted due to inversion symmetry breaking.}
\label{fig:so-proj-band}
\end{figure}
\begin{figure*}[htbp!]
\centering
\includegraphics[width=0.45\textwidth,clip=true]{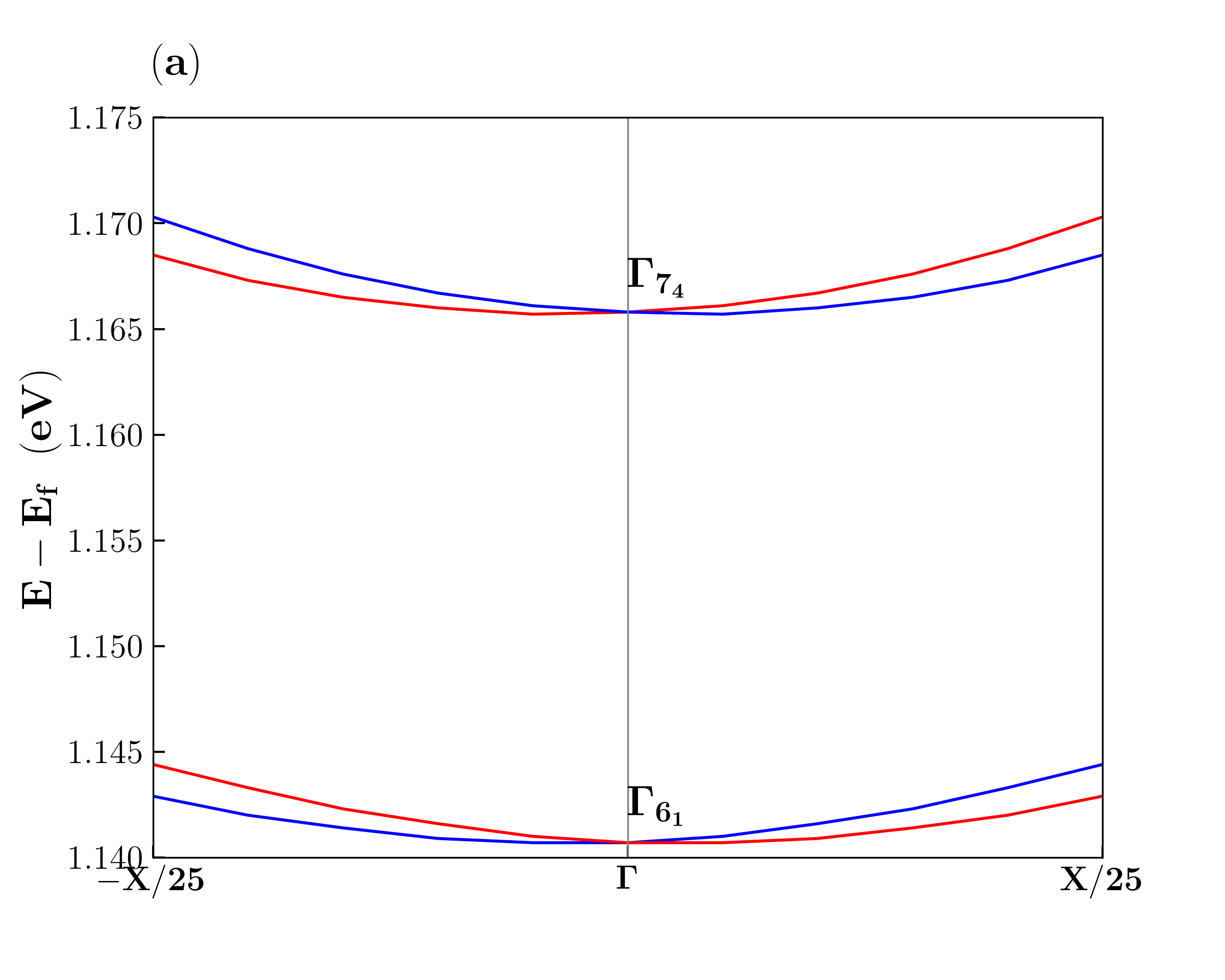}
\includegraphics[width=0.45\textwidth,clip=true]{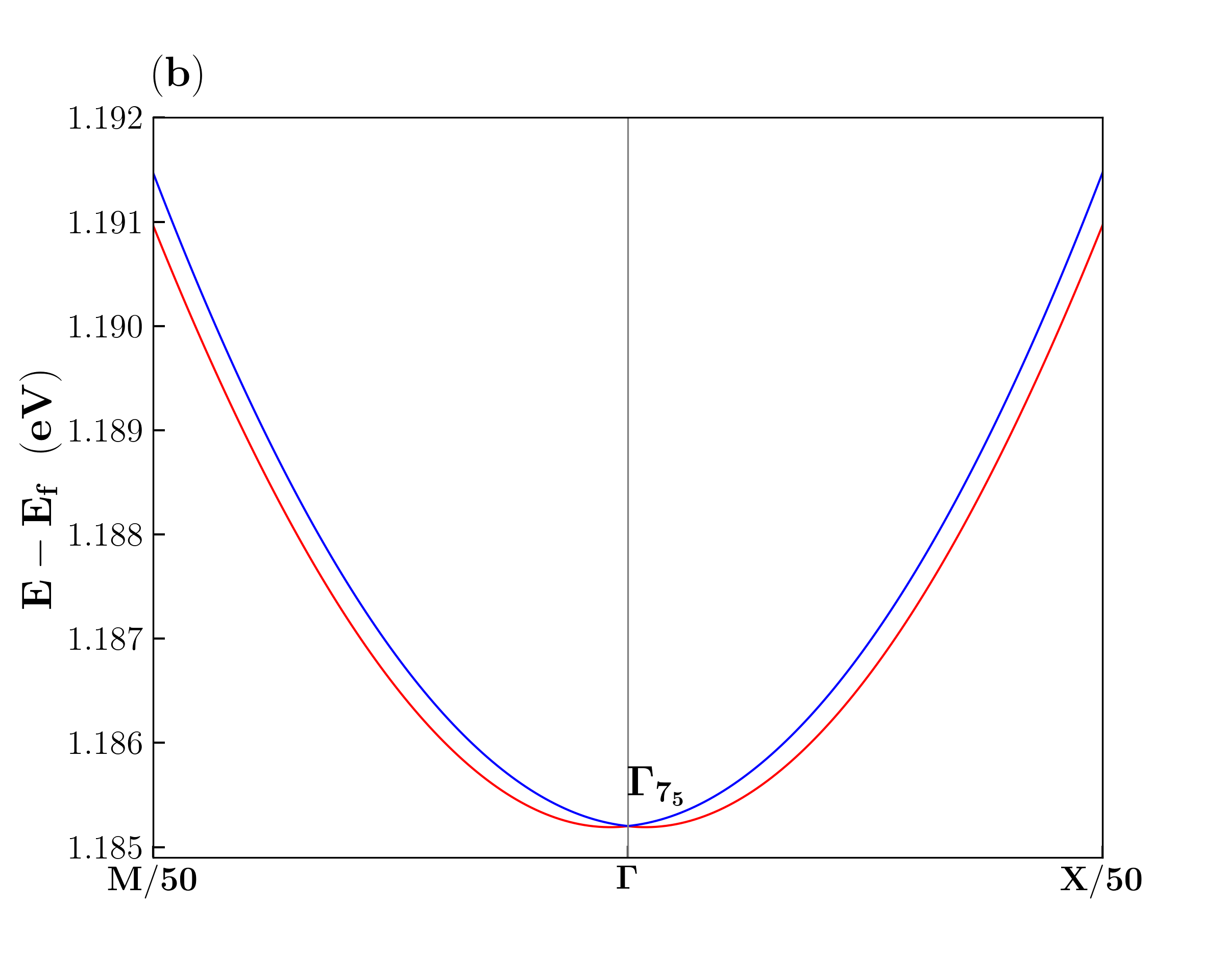}
\caption{Spin-splitting band structure of Ti-$\rm t_{2g}$ orbital originating from $\rm {TiO_2}-I$ layer. (a) present the spin-splitted $\Gamma_{6_1}$ and $\Gamma_{7_4}$ bands along $\rm{-X}$--${\Gamma}$--$\rm{X}$ direction with the projection of electron spins towards $\rm \hat{y}$. (b) represents $\Gamma_{7_5}$ band along $\rm{M}$--${\Gamma}$--$\rm{X}$--$\rm{M}$. Red and blue coloured lines represent positive and negative spin polarization along $\rm \hat{y}$ direction. }
\label{fig:delta}
\end{figure*}
Moreover, Ti-3d orbitals experience a further splitting in the presence of SOC. In our DFT calculations, we have established that the wave vectors ($\vec{k}$) consistently lie in the $\rm k_x$--$\rm k_y$ plane, aligning along the (1 0 0) direction between $\Gamma$ and X, and the (1 1 0) direction along the $\Gamma$--$\rm{M}$ path.
The spin component is zero in the z [0 0 1] direction, corresponding to the polar axis normal to the surface, which is a characteristic feature of the Rashba effect. A comprehensive band structure depicting different spin polarizations is presented in Appendix\ref{bs}. Around the $\Gamma$, $\rm Ti-3d_{yz/zx}$ bands originating from different layers coincide, making them indistinguishable layer by layer.
In the presence of SOC, the true eigenstates of the system are combinations of the original $d-$orbitals~\cite{khalsa-orbital-mixing}, and hence it is imperative to identify the 
bands in terms of its total angular momentum quantum number $j$ and the corresponding $m_j$ values. In bulk $\rm{SrTiO_3}$, around the $\Gamma$ point, the original $e_g$ and $t_{2g}$ bands remain well 
separated even in the presence of SOC. SOC split the $t_{2g}$ band to a twofold degenerate $\Gamma_7$ and a fourfold degenerate $\Gamma_8$ level~\cite{khalsa-orbital-mixing}. This fourfold 
$\Gamma_8$ level is identified with $\ket{3/2,\pm3/2}$ and $\ket{3/2,\pm1/2}$~\cite{Fasolino,shanavas-2016-sto}. The $\Gamma_8$ level further splits to $\Gamma_6$ and $\Gamma_7$ under tetragonal 
distortion~\cite{khalsa-orbital-mixing}. Surprisingly, however, in our system, we have observed from our data that there is mixing of the original $e_g$ and $t_{2g}$ bands in the presence of SOC and tetragonal 
distortion. In Fig.\ref{fig:so-proj-band}, we present fat bands illustrating different angular momentum components. Our analysis reveals a significant mixing of orbitals within each band. For instance, in the $\Gamma_{7_1}$ band, we observe the presence of $\ket{\frac{5}{2},{\pm\frac{5}{2}}}$, $\ket{\frac{5}{2},{\pm\frac{3}{2}}}$, and $\ket{\frac{3}{2},{\pm\frac{3}{2}}}$. Similarly, in the $\Gamma_{6_1}$ band is found to be a combination of $\ket{\frac{5}{2},{\pm\frac{1}{2}}}$ and $\ket{\frac{3}{2},{\pm\frac{1}{2}}}$. The combination of $\ket{j,m_j}$ for all the bands designated in Fig.~\ref{fig:so-proj-band} are presented in Table.\ref{tab_ang}. Interestingly, this kind of interband coupling has not been reported earlier in $\rm SrTiO_3$. However, such interband mixing is not unusual. For example, in reference~\cite{MinsungKim}, a significant $\rm e_g$ and $\rm t_{2g}$ interband coupling has been found in a tantalate monolayer on $\rm BaHfO_3$, which arises due to strong confinement that breaks the inversion symmetry maximally. 
More recently it has been pointed out that such an interband coupling is crucial to explain the RSS in $\rm KTaO_3$ heterostructure~\cite{varotto2022direct}.
\begin{table}
\caption{This table represents the role of total angular momentum quantum number $j$ and the corresponding $m_j$ values present in each band designated in Fig.~\ref{fig:so-proj-band}.}
\begin{tabular}{c c}
	\hline
		Band level & $\ket{j,m_j}$ \\
			\hline\hline
			&\\
$\Gamma_{7_1}$, $\Gamma_{7_2}$, $\Gamma_{7_3}$, $\Gamma_{7_5}$, $\Gamma_{7_6}$, $\Gamma_{7_7}$ & $\ket{\frac{5}{2},{\pm\frac{5}{2}}}$, $\ket{\frac{5}{2},{\pm\frac{3}{2}}}$, $\ket{\frac{3}{2},{\pm\frac{3}{2}}}$ \\
&  \\
		
$\Gamma_{7_4}$, $\Gamma_{7_8}$, $\Gamma_{7_9}$, $\Gamma_{7_{10}}$, $\Gamma_{7_{11}}$, $\Gamma_{7_{12}}$ & $\ket{\frac{5}{2},{\pm\frac{3}{2}}}$, $\ket{\frac{3}{2},{\pm\frac{3}{2}}}$ \\
& \\

$\Gamma_{6_1}$, $\Gamma_{6_2}$, $\Gamma_{6_3}$, $\Gamma_{6_4}$, $\Gamma_{6_5}$, $\Gamma_{6_6}$ & $\ket{\frac{5}{2},{\pm\frac{1}{2}}}$, $\ket{\frac{3}{2},{\pm\frac{1}{2}}}$\\
&\\
\hline
\end{tabular}
\label{tab_ang}
\end{table}
\subsection{RSS and spin texture}\label{rs}
Next, we proceed to analyse the splitting of band pairs in this structure.
To quantify the band splitting and comprehend the impact of symmetry on the spin texture, we employ the $\Vec{k}.\Vec{p}$ Hamiltonian associated with the structure. The introduction of a sandwich structure alters the point group symmetry of $\rm {SrTiO_3}$ to $c_{4v}$. The high symmetry points of the 2D slab structure are $\Gamma$ (0, 0, 0), X (0.5, 0, 0), and M (0.5, 0.5, 0). In Fig.\ref{fig:so-proj-band}, the levels denoted as $\Gamma_{7_1}$ and $\Gamma_{7_5}$ originate from $\rm TiO_2$-V and $\rm TiO_2$-I layers, respectively, while $\Gamma_{6_1}$ and $\Gamma_{7_4}$ are present in all $\rm TiO_2$ layers. A zoomed-in view of the $\Gamma_{6_1}$, $\Gamma_{7_4}$, and $\Gamma_{7_5}$ bands is presented in Fig.~\ref{fig:delta}(a) and \ref{fig:delta}(b).
To study the RSS in these bands, an effective two-band Hamiltonian derived from the $ \Vec{k}.\Vec{p}$ model is presented here to characterize the splitted bands at $\Gamma$ point. The Hamiltonian for the $c_{4v}$ point group~\cite{hamiltonian-c4v-vajna,shanavas-2016-sto,arras-pbtio3-2019} is written up to the third order:
\begin{equation}
\begin{split}
H_{c_{4v}} = \gamma ({k_x}^{2}+{k_y}^{2}) + \beta {k_z}^{2} +\alpha_{R_1} ({k_x}{\sigma_{y}}-{k_y}{\sigma_{x}}) \\
+ \alpha_{R_2}{k_x}{k_y}({k_y}{\sigma_{y}}-{k_x}{\sigma_{x}})+ \alpha_{R_3}({{k_x}^{3}}{\sigma_{y}}-{{k_y}^{3}}{\sigma_{x}})
\end{split}
\label{hamiltonian}
\end{equation}
The parameter $\gamma$ is related to the effective mass ${m}^*$ by the $\lvert\gamma\lvert = \frac{\hbar^2}{2m^*}$ of the associated bands, $\alpha_{R_1} ({k_x}{\sigma_{y}}-{k_y}{\sigma_{x}})$ is the linear RSO interaction term, whereas $\alpha_{R_2}$ and $\alpha_{R_3}$ are the coefficients of cubic RSO interaction, with the cubic RSO interaction term $\alpha_{R_2}({k_x}{k_y}^{2}{\sigma_{y}}-{k_x}^{2}{k_y}{\sigma_{x}})$ and $ \alpha_{R_3}({{k_x}^{3}}{\sigma_{y}}-{{k_y}^{3}}{\sigma_{x}})$. $k_i$ (i = x, y, z) are the components of the wave vector $\vec{k}$ close to the HS points, and $\sigma_i$ are the Pauli spin matrices. 

To provide a basic characterization of the energy spin splitting, irrespective of the point group, one can model the difference in energy between the two bands of opposite spin, denoted as $\rm\Delta{E_{\pm}}$, as a third-order polynomial in terms of the wave vector $\vec{k}$, specifically, $\rm \Delta{E_{\pm}(k)}= ak+bk^3$. 
This polynomial also follows the fitting criteria obtained from the eigenvalue Eqn.\ref{hamiltonian}.
The detailed derivation of the eigenvalues for different k-paths is given in Appendix~\ref{effective hamiltonian}. For the energy spin splitting along the $\Gamma$--$\rm{X}$ path, the coefficient $\rm b$ is determined as $2\alpha_{R_3}$, while along the $\Gamma$--$\rm{M}$ direction, $\rm b=\alpha_{R_2}+\alpha_{R_3}$. In both directions along the k-path, $\rm a$ is consistently equal to $2\alpha_{R_1}$. These findings collectively contribute to generating contour and orbital spin textures around the $\Gamma$ point in the $\rm k_x$--$\rm k_y$ plane. Next, we will apply this method to analyze the Rashba spin splitting (RSS) and the spin texture in the $\Gamma_{6_1}$, $\Gamma_{7_1}$, $\Gamma_{7_4}$, and $\Gamma_{7_5}$ bands.
To determine the contour for the particular pair of spin-splitting bands, we fit our DFT data numerically with Hamiltonian presented in eqn.~\ref{hamiltonian}. The k-limit is important in determining the spin texture and concentric iso-energy contour.
\begin{figure*}[htbp!]
\centering
\includegraphics[scale=0.35,trim={0.4cm 0cm 0.2cm 0.2cm},clip=true]{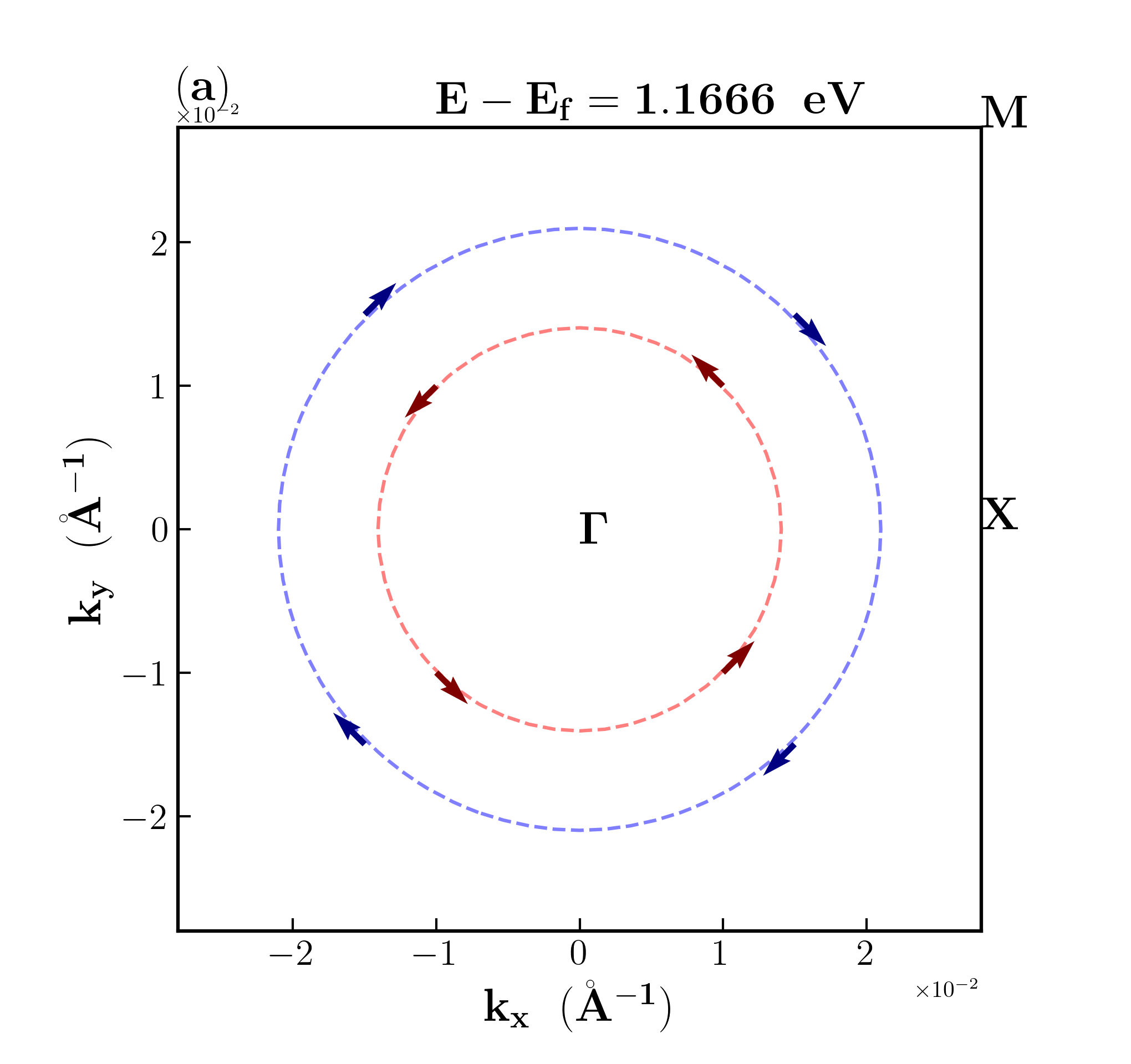}\hspace{-4.5mm}
\includegraphics[scale=0.35,trim={0.2cm 0cm 0.2cm 0.2cm},clip=true]{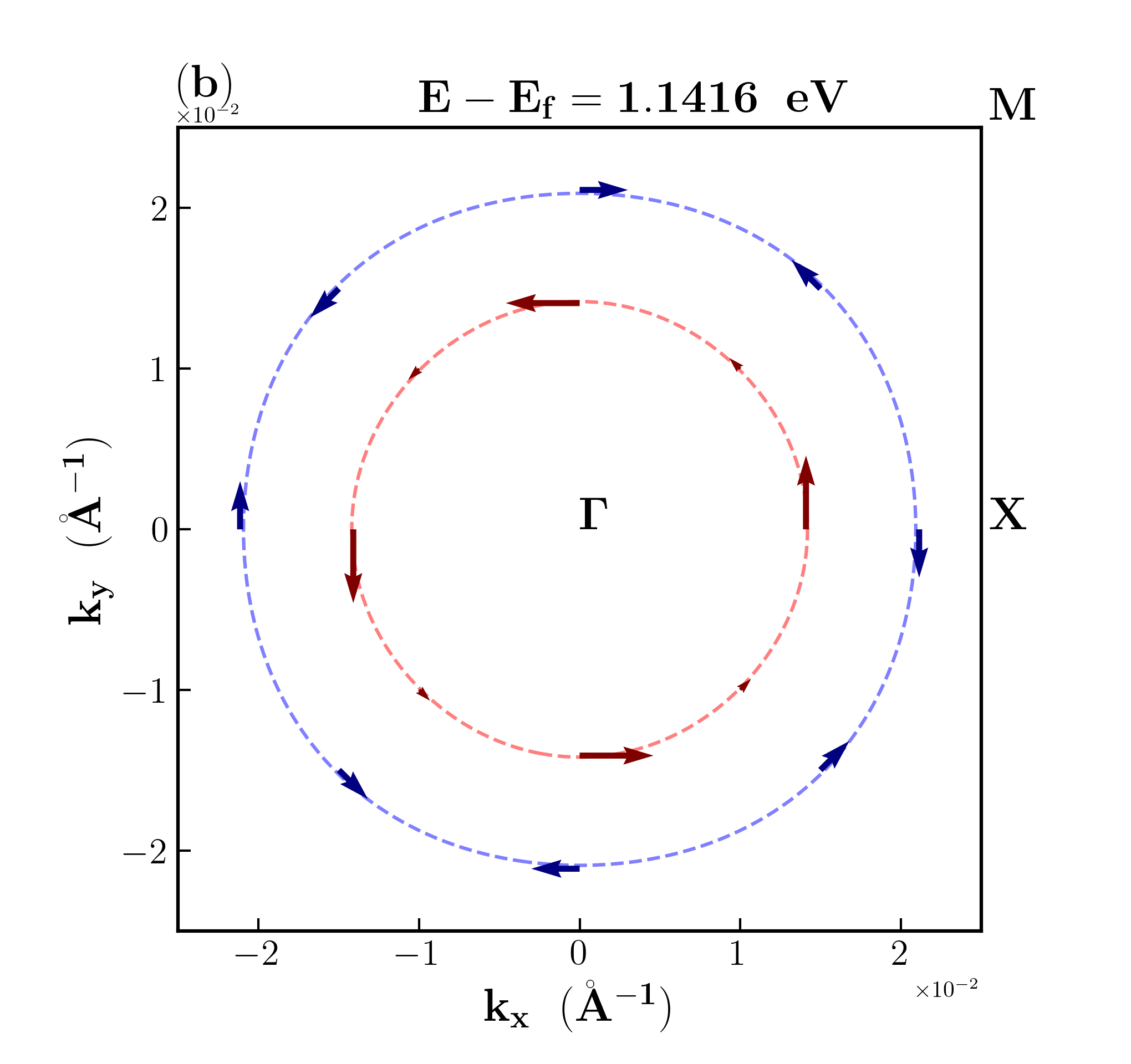}\hspace{-4.5mm}
\includegraphics[scale=0.35,trim={0.2cm 0cm 0.4cm 0.2cm},clip=true]{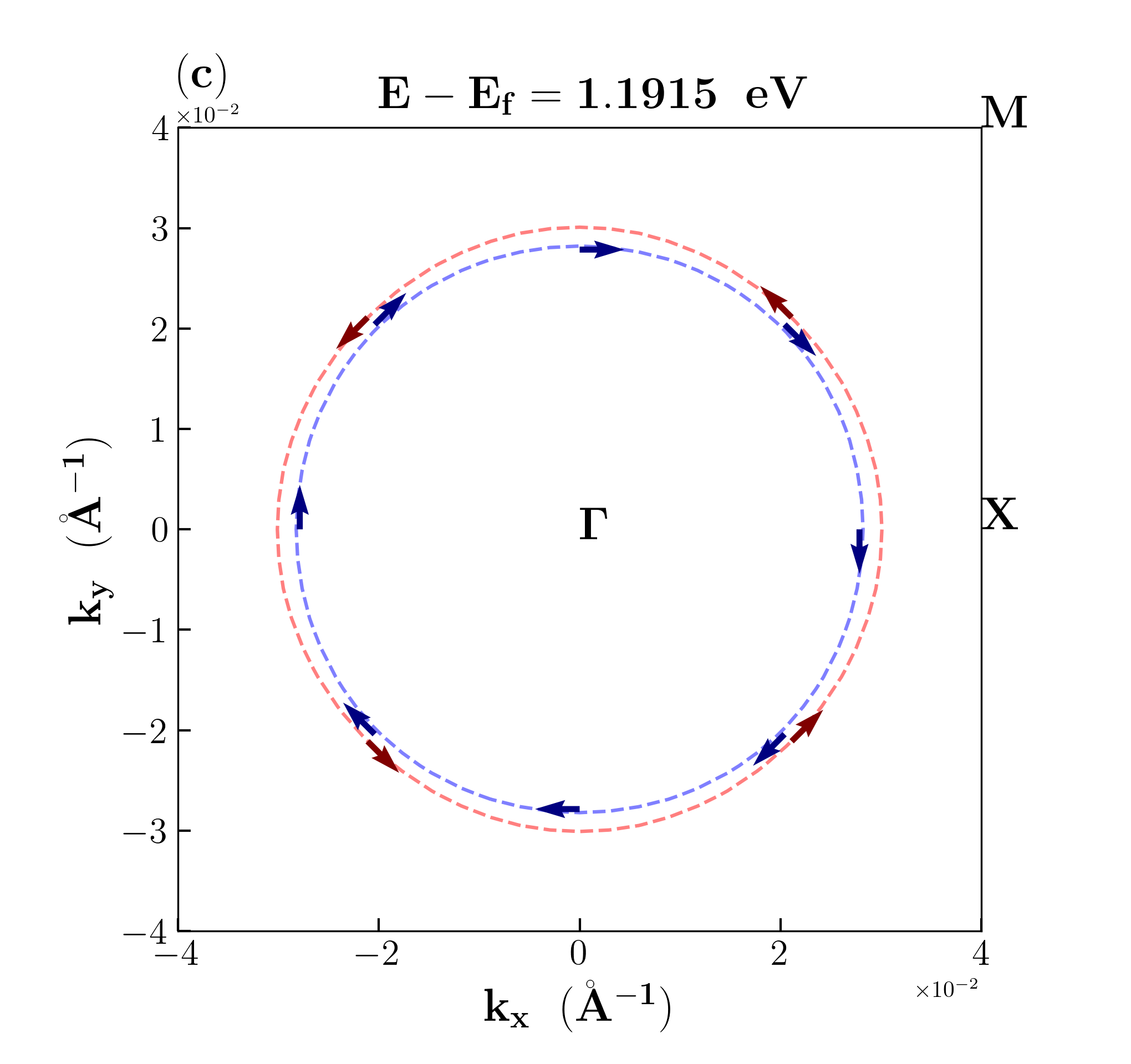}
\caption{The spin texture derived from DFT data are projected on $\rm k_x$--$\rm k_y$ plane for, (a)$\Gamma_{7_4}$ bands with energy contour at $\rm E-E_{f}= 1.1666~eV$, (b) $\Gamma_{6_1}$ bands with energy contour at $\rm E-E_{f}= 1.1416~eV$, and (c)$\Gamma_{7_5}$ bands with energy contour at $\rm E-E_{f}= 1.1915~eV$. The contours are derived by using the Hamiltonian presented in Eqn.~\ref{hamiltonian} along with a reasonable $\rm k_{limit}$.}
\label{fig:dft-spin}
\end{figure*}
\begin{figure*}[htbp!]
\centering
\includegraphics[scale=0.3,trim={0.09cm 0cm 0.2cm 0.2cm},clip=true]{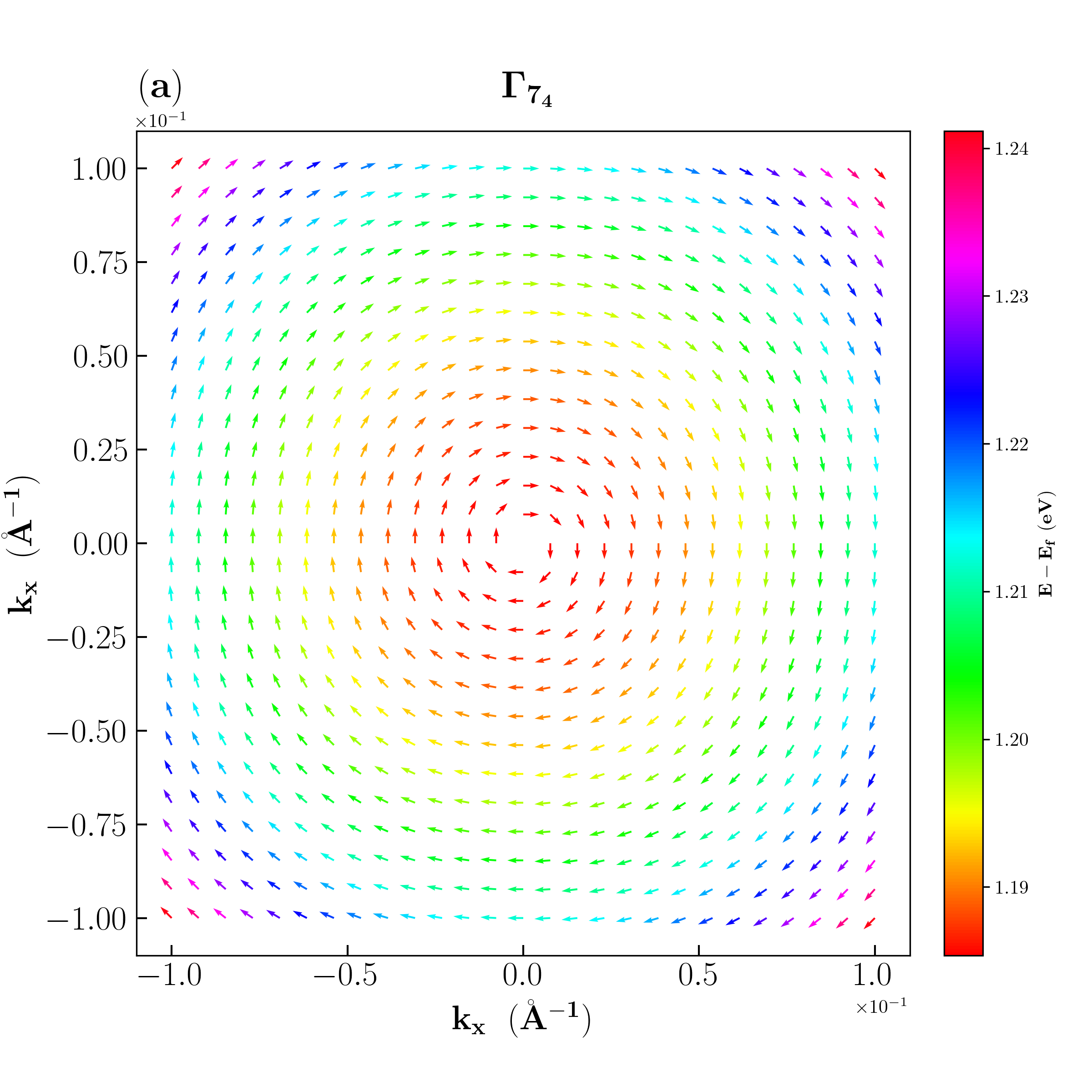}\hspace{-1mm}
\includegraphics[scale=0.3,trim={0.69cm 0cm 0.2cm 0.2cm},clip=true]{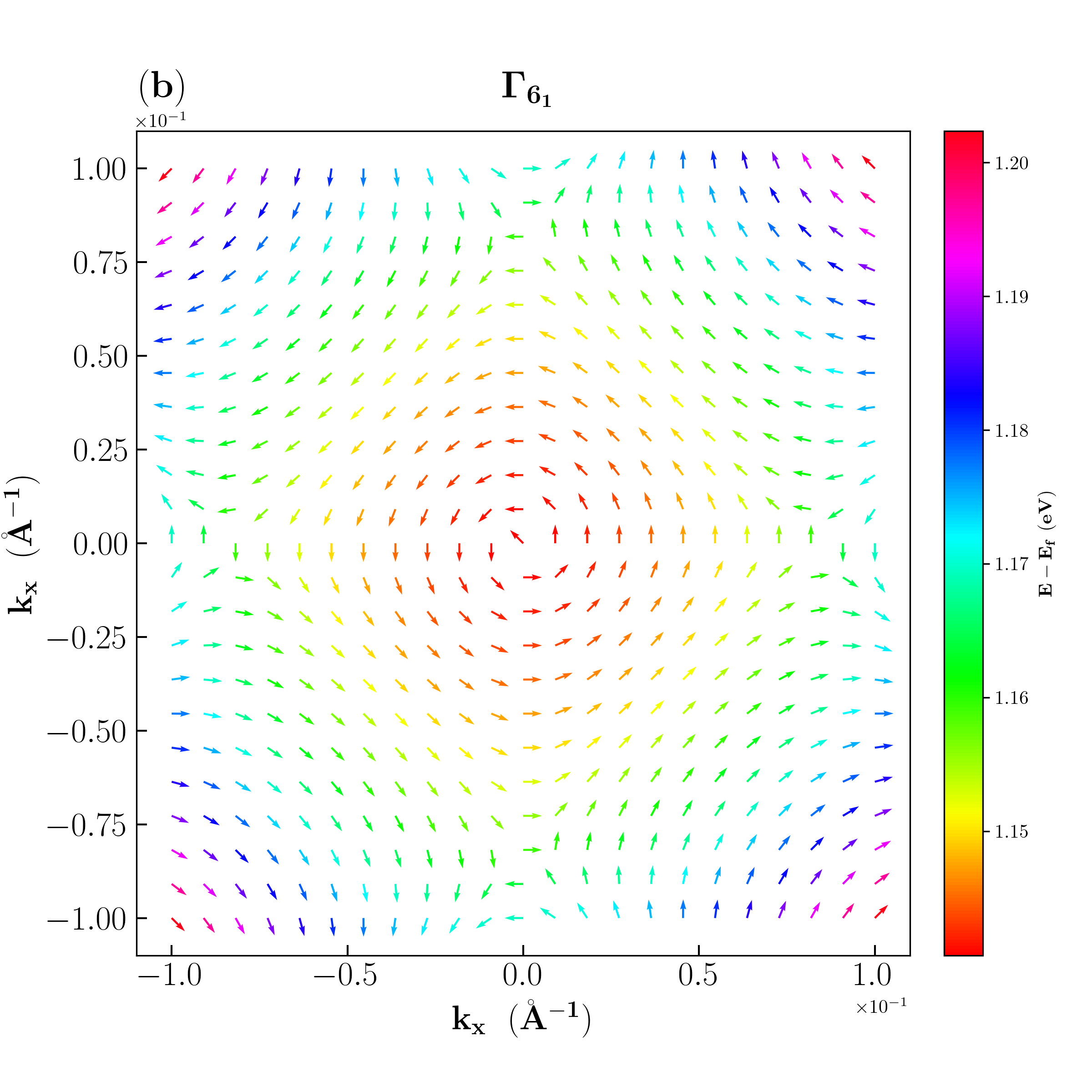}\hspace{-1mm}
\includegraphics[scale=0.3,trim={1.1cm 0cm 0.4cm 0.2cm},clip=true]{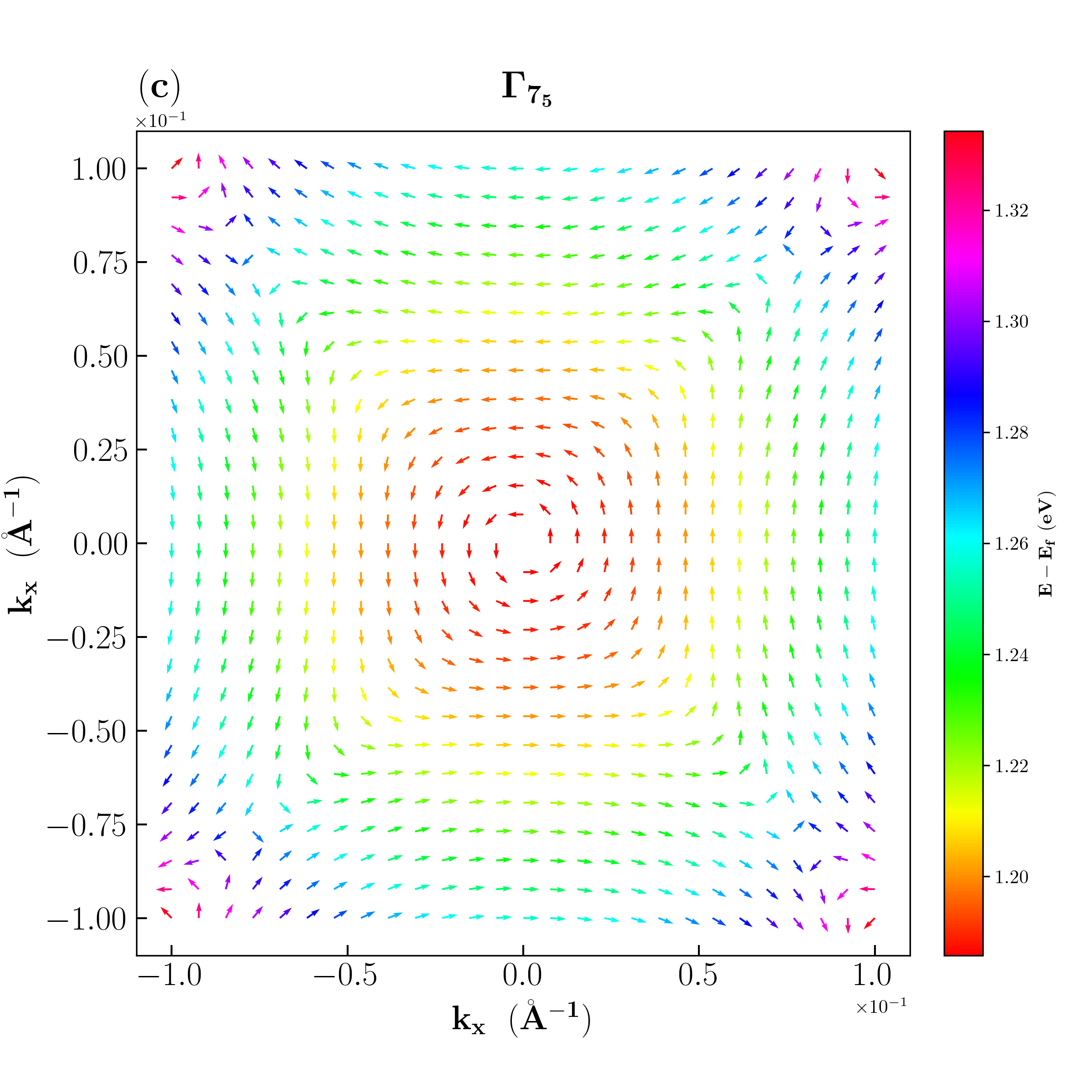}
\caption{One band spin projection in the $\rm k_x$--$\rm k_y$ plane derived from the Hamiltonian (Eqn.~\ref{hamiltonian}).(a) and (c) present the outer band projection. (b) presents the inner band projection. The orientation of the spins are taken same as that of the DFT spin orientation (Fig.~\ref{fig:dft-spin}). The length of the arrows are not the real length of arrows presented in DFT spin texture.}
\label{fig:spin-texture}
\end{figure*}
\begin{table}[b]  
\caption{The table represents the effective mass $ m{^*} = \frac{\hbar^2}{2\lvert\gamma\lvert}$, linear and cubic Rashba spin-splitting parameter $\rm \alpha_{R_1}$, $\rm \alpha_{R_3}$ \& $\rm \alpha_{\Tilde{R3}}$ for different band levels along $\Gamma$-X and $\Gamma$-M path obtained from the two-band fitting with polynomial 
$\rm \Delta{E_{\pm}(k)}= ak+bk^3$.}
\centering  
\begin{tabular}{c c c c c c c c } 
\hline   
Band &  k-path & $\rm E-E_f$    &  $m^*$ &  $\rm \alpha_{R_1}$ &  $\rm \alpha_{R_3}$ & $\rm \alpha_{\Tilde{R_3}}$\\
Level& direction &(eV)     &($m_0$)  &   (eV\AA)   & (eV\AA$^3$)   & (eV\AA$^3$)\\
\hline\hline   
$\Gamma_{7_1}$ &$\Gamma_{\rightarrow X}$ & 0.9035  & 0.5& 3.24$\times10^{-3}$  &0.43 &\\

&$\Gamma_{\rightarrow M}$& 0.9035    & 0.5& 4.12$\times10^{-3}$ & &0.33 \\

$ \Gamma_{6_1}$ & $\Gamma_{\rightarrow X}$& 1.1407  & 1.46 & 2.55$\times10^{-2}$ & -2.16 &\\

$ \Gamma_{7_4}$ & $\Gamma_{\rightarrow X}$& 1.1658 & 1.12 & 2.66$\times10^{-2}$ & 1.27 &\\

&$\Gamma_{\rightarrow M}$  &  1.1658 & 1.23 & 2.67$\times10^{-2}$ & &0.13  \\

$ \Gamma_{7_5}$ & $\Gamma_{\rightarrow X}$& 1.1852 & 0.5 & 1.11$\times10^{-2}$ & -3.33 &\\

&$\Gamma_{\rightarrow M}$& 1.1852    & 0.51& 1.77$\times10^{-2}$ & & -1.34 \\
\hline
\end{tabular}  
\label{tab1}
\end{table}
We commence our analysis with the $\Gamma_{6_1}$ band pairs of $\rm TiO_2$-V atomic layer. To understand the Rashba spin-splitting. Along $\Gamma$--$\rm X$, the effective mass of the $\Gamma_{6_1}$ band is determined to be 1.46 $m_0$. The linear and cubic Rashba spin-orbit (RSO) coefficients are found to be $\rm \alpha_{R_1}$= 0.025 eV\AA~and $\rm \alpha_{R_3}$=-2.16 eV\AA$^{3}$ within a k-limit of 0.046 \AA$^{-1}$. However, along the $\Gamma$--$\rm M$ direction, spin-splitting is observed, though it may not strictly adhere to the Rashba-type. This characteristic is further confirmed by the iso-energy spin texture with energy contour in the $\Gamma$--$\rm X$--$\rm M$ plane presented in Fig.~\ref{fig:dft-spin}(b), where the two bands exhibit different spin-topologies, potentially signifying anisotropic RSOC in the system.
The disparity in spin length along the inner band for the entire path suggests a potential helical spin texture. This phenomenon, although not uncommon, has been observed previously in the $\rm KTaO_3$/$\rm LaAlO_3$ heterostructure, as reported in earlier first-principle studies~\cite{bhattacharya2023evidence}.
In the atomic layer $\rm TiO_2$-V, the next band pair under consideration is the $\Gamma_{7_4}$~\ref{fig:delta}(a). The effective mass of the band along $\Gamma$--$\rm X$ is 1.12 $m_0$, and along $\Gamma$--$\rm M$ is 1.23 $m_0$. These bands also display an iso-energy spin texture, featuring two concentric circles with tangent spins rotating in inverse directions, indicative of a cubic Rashba spin-splitting. This is corroborated by our fits of the $\Gamma_{7_4}$ bands around the $\Gamma$ point, yielding an isotropic cubic-Rashba parameter of 1.27 eV\AA$^{3}$ along $\Gamma$--$\rm X$ and 0.13 eV\AA$^{3}$ along $\Gamma$--$\rm M$. The linear RSO coefficient for this band is determined to be 0.027 eV\AA. The prevailing cubic Rashba spin splitting is further exemplified by the DFT spin texture presented in \ref{fig:dft-spin}(a).
The Rashba spin-splitting of the nearest bands to the Fermi energy is denoted as $\Gamma_{7_1}$. The effective mass of the band is 0.5 $m_0$ around $\Gamma$-point. The coefficient of linear spin-splitting for$\Gamma_{7_1}$ band pair is found to be 3.235$\times10^{-3}$ eV\AA~and 4.12$\times10^{-3}$ eV\AA~along $\Gamma$--$\rm X$ and $\Gamma$--$\rm M$, respectively. In this $\Gamma_{7_1}$ level, the cubic Rashba is more prominent than the linear, and the cubic RSO coefficients are found to be 0.43 eV\AA$^{3}$ and 0.33 eV\AA$^{3}$ along $\Gamma$--$\rm X$ and $\Gamma$--$\rm M$, respectively. The DFT spin texture is presented in Fig.\ref{fig:dft-spin}(c).
Finally, we discuss the band $\Gamma_{7_5}$, originating from $\rm TiO_2$-I layer, which is the closest to the n-type interface. The effective mass of the band is 0.5 $m_0$. Along $\Gamma$--$\rm X$, it exhibits predominantly cubic Rashba spin-splitting with -3.33 eV\AA$^{3}$ up to 0.15 \AA$^{-1}$, while the linear splitting is 1.11$\times10^{-2}$, retaining linearity in the closest vicinity of the $\Gamma$ point, i.e., up to 0.03 \AA$^{-1}$. Along $\Gamma$--$\rm M$, the linear coefficient of RSOC is 1.77$\times10^{-2}$, and the cubic Rashba coefficient is determined to be -1.34 eV\AA$^{3}$. The summarized results are presented in Table.\ref{tab1}. For the remaining bands, we do not observe any significant Rashba spin splitting around $\Gamma$. However, the $\Gamma_{7_3}$ band pairs exhibit splitting primarily around the band crossing region, and the Rashba spin splitting is distinctly cubic with $\rm \alpha_{R_3}=$ 1.86 eV\AA$^{3}$ along $\Gamma$--$\rm X$ and $\rm \alpha_{\Tilde{R_3}}=$ 1.2 eV\AA$^{3}$ along $\Gamma$--$\rm M$. The other $\Gamma_{6}$ levels, located at higher energy regions, do not display Rashba-like spin splitting. Furthermore, no linear Rashba spin splitting is evident in the spin-splitting of the Ti bands.

It is already known that the spin orientation changes as a function of the wave vector $\Vec{k}$ together with the modification of the orbital contributions of the interface states. 
We derive the projected spin components ${\langle S_x \rangle}$ and ${\langle S_y \rangle}$ from our density functional theory (DFT) results to explore the potential presence of spin arrangements for a single band of the spin-splitted pair. Precise determination of the RSS is a cumbersome task within specific $\rm k_{limit}$, however, the orientation of the spin can be determined properly from the DFT spin texture. However, the length of the spin presented in the one band spin texture are not the real data.  Subsequently, we depict the projected spin vectors for a single band as ${\langle S_x \rangle}\hat{i}+{\langle S_y \rangle}\hat{j}$ in the $\rm{k_x}$--$\rm{k_y}$ plane, within a suitable energy range. These figures are presented for one band spin texture in Figs.~\ref{fig:spin-texture}(a), (b), and (c).
Fig.~\ref{fig:spin-texture}(a) represents the one band spin texture for $\Gamma_{7_4}$ band which is pure cubic type RSS around the $\Gamma$ point in the $\rm X$--$\Gamma$--$\rm M$ plane. On the contrary, Fig.~\ref{fig:spin-texture}(b) portrays the anisotropy of the $\Gamma_{6_1}$ band pair. In Fig.~\ref{fig:spin-texture}(c), the spins are oriented perpendicular to the $\Gamma$--$\rm M$ line, which indicates the absence of $\rm \alpha_{R_2}$ in the Hamiltonian of the  $\Gamma_{7_5}$ bands.
\section{Conclusion}
In conclusion, our investigation into Rashba spin-orbit coupling in the LAO/STO/LAO sandwich heterostructure involves the construction of a model featuring a thin STO slab with 5.5 unit cells. Through DFT-based first-principle calculations, we delve into the intricate role of orbitals within different layers of the STO slab. 
We find a presence of significant interorbital mixing of $\rm t_{2g}$ and $\rm e_g$ orbital in the STO slab. We report this phenomenon for the first time in STO-based 2D structure. The identification of this mixing in sandwich structure is pivotal, providing valuable insights for future researchers in modelling and understanding the complexities of the system.
Moreover, in this study, we encompass the spin texture for splitted bands to elucidate the nature of Rashba spin splitting in the system, and reveal a predominant cubic Rashba spin-orbit coupling (RSOC). 
This research not only enriches the understanding of the specific heterostructure but also establishes a foundation for prospective advancements in comprehending Rashba spin-orbit coupling in complex oxide heterostructures.
%
\appendix
\begin{figure*}[h]
\centering
\includegraphics[scale=0.32,trim={0.1cm 1.4cm 1cm 1cm},clip=true]{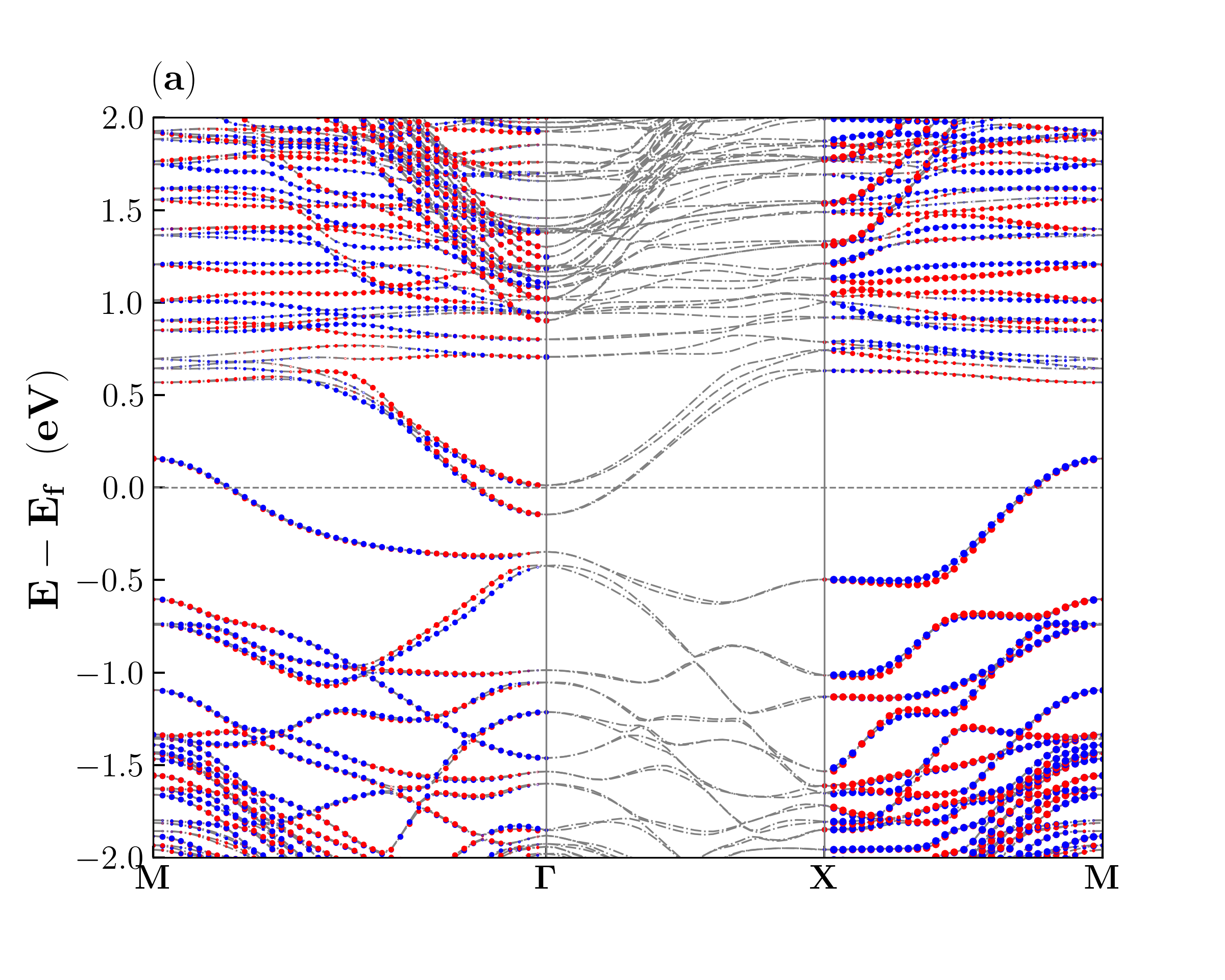}\hspace{-0.5mm}
\includegraphics[scale=0.32,trim={1.2cm 1.4cm 1cm 1cm},clip=true]{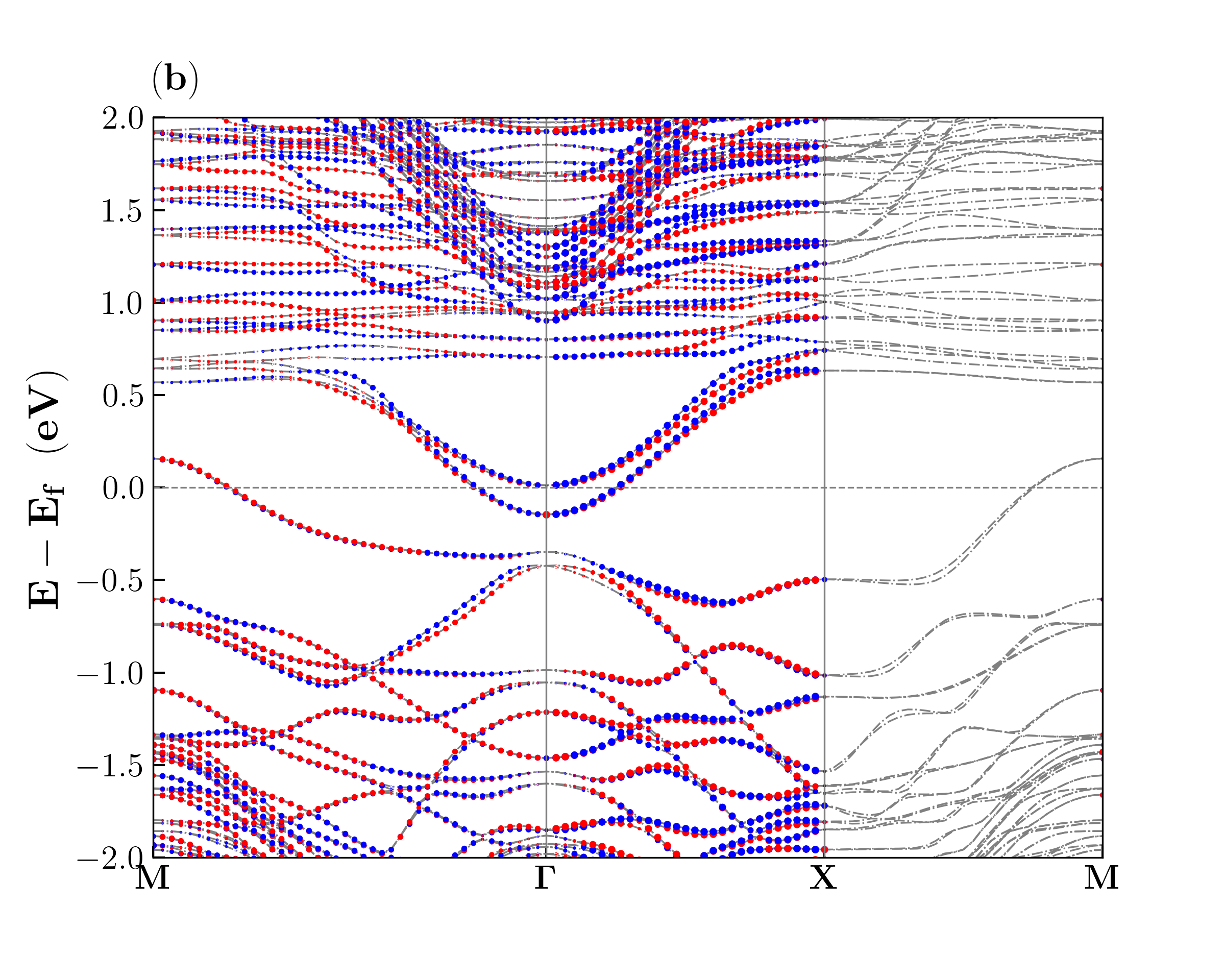}\hspace{-0.5mm}
\includegraphics[scale=0.32,trim={1.2cm 1.4cm 1cm 1cm},clip=true]{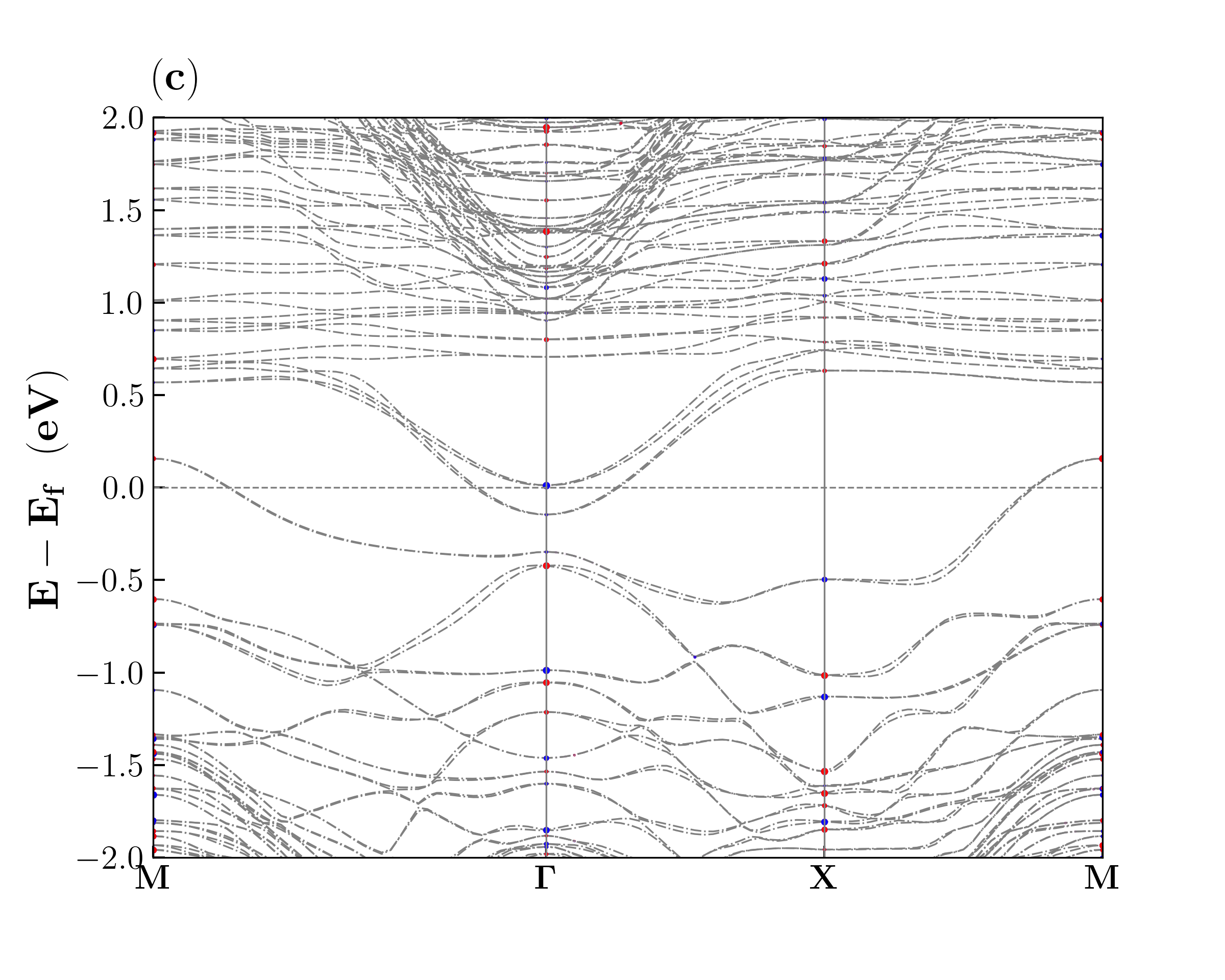}
\caption{Band structures of the LAO/STO/LAO sandwich model along $\rm{M}$--${\Gamma}$--$\rm{X}$--$\rm{M}$ calculated with the spin-orbit interaction. Red and blue coloured circles represent up and down spin polarization. The circle size represents the weight of the spin. (a), (b) and (c) represent the $\rm{s_x}$, $\rm{s_y}$ and $\rm{s_z}$ spin-polarization projected along $\rm{M}$--${\Gamma}$--$\rm{X}$--$\rm{M}$.}
\label{fig:soc-1}
\end{figure*}
\section{SOC included band structure}\label{bs}
This section provides the band structure for the entire structure (Fig.~\ref{fig:pn-struct}) along $\rm{M}$--${\Gamma}$--$\rm{X}$--$\rm{M}$. The band structure presented in Fig.~\ref{fig:soc-1} highlights the unique characteristics of the Rashba effect within the system. Fig.~\ref{fig:soc-1}(a), (b) and (c) represent the band structure with spin-polarization along $\hat{x}$, $\hat{y}$ and $\hat{z}$. Specifically, the spins exhibit polarization perpendicular to the wave vector. The red and blue solid circles represent the positive and negative spin, whereas the size of the circles represent the weight of the spin associated with the bands. Along $\hat{z}$, the spin weight are present only at the high symmetry points, which is a characteristics of the RSOC in the system. The complete system, including the capping LAO layers, exhibits metallic states. These metallic states may be attributed to either oxygen (O) or lanthanum (La) in the LAO layers, as depicted in Figure \ref{fig:lrdos}.
\section{Effective $\bold{k.p}$ Hamiltonian}\label{effective hamiltonian}
The strength of the RSO interaction can be estimated by solving Eq.~\ref{hamiltonian} separately for $\Gamma-\rm{X}$ 
and $\Gamma-\rm{M}$ high symmetry path, represented in the Table~\ref{tab1}. It is evident that the Hamiltonian of Eq.~\ref{hamiltonian} only exhibits terms that rely on $\sigma_{x}$ and $\sigma_{y}$, as per symmetry concerns. We confirm from our DFT findings that the spin polarisation contribution is zero along the z [0 0 1] axis for every $\vec {k}$ vector.

For $\Gamma-\rm{X}$ high symmetry path, $\rm k_{y}=0$. 
Imposing this condition, we solve Eq.~\ref{hamiltonian} to obtain the following two eigenvalue equations are given by,
\begin{equation*}
\epsilon_{1} = \gamma k_{x}^{2} + \alpha_{R_1} k_{x} + \alpha_{R_3} k_{x}^{3}
\end{equation*}
\begin{equation*}
\epsilon_{2} = \gamma k_{x}^{2} - \alpha_{R_1} k_{x} - \alpha_{R_3} k_{x}^{3}.
\end{equation*}
The amount of momentum dependent spin-splitting $\Delta_{RX}$ is given by,
\begin{equation}
\Delta_{RX} = \epsilon_{1} - \epsilon_{2} = 2 \alpha_{R_1} k_{x} + 2 \alpha_{R_3} k_{x}^{3}
\label{fit1}
\end{equation}

For $\Gamma-\rm{M}$ high symmetry path, $k_x=k_y$, and $k_{||}=\sqrt{k_x^2+k_y^2}$.
By imposing these conditions, solutions for Eq.~\ref{hamiltonian} give two eigenvalue equations as follows,
\begin{equation*}
\epsilon_{1}^{'} = \gamma k_{||}^{2} + \alpha_{R_1} k_{||} + \Big(\frac{\alpha_{R_2}+\alpha_{R_3}}{2}\Big) k_{||}^{3}
\end{equation*}
\begin{equation*}
\epsilon_{2}^{'} = \gamma k_{||}^{2} - \alpha_{R_1} k_{||} - \Big(\frac{\alpha_{R_2}+\alpha_{R_3}}{2}\Big) k_{||}^{3}.
\end{equation*}
The amount of momentum dependent spin-splitting $\Delta_{RM}$ is given by,
\begin{eqnarray}
\Delta_{RM} &= &\epsilon_{1}^{'} - \epsilon_{2}^{'} \nonumber\\
&=& 2 \alpha_{R_1} k_{||} +(\alpha_{R_2}+\alpha_{R_3}) k_{||}^{3} \nonumber\\
& = &2 \alpha_{R_1} k_{||} +(\Tilde{\alpha_{R_3}}) k_{||}^{3}  
\label{fit2}
\end{eqnarray}
where, $\Tilde{\alpha_{R_3}} = (\alpha_{R_2}+\alpha_{R_3})$ is the modified cubic RSO interaction strength.\\
The linear and cubic coupling strengths are obtained by fitting Eq.~\ref{fit1} and Eq.~\ref{fit2} to the DFT data. All the RSO interaction strengths are presented in Table\ref{tab1}.
The eqns.~\ref{fit1} and ~\ref{fit2} can be written in the form of polynomial $\rm ak+bk^3$.
\end{document}